\documentclass[
aps,%
12pt,%
final,%
notitlepage,%
oneside,%
onecolumn,%
nobibnotes,%
nofootinbib,%
superscriptaddress,%
noshowpacs,%
centertags]%
{revtex4}

\usepackage{graphicx}
\usepackage{morefloats}
\usepackage{epstopdf}
\usepackage{amssymb,amsmath}

\begin{document}

\centerline{\bf Tricolor Technique for Visualization of Spatial Variations}
\centerline{\bf of Polydisperse Dust in Gas-Dust Flows}

{\it Vitaly V. Korolev $^{1}$, Mikhail A. Bezborodov $^{1}$, Ilya G. Kovalenko $^{1,*}$, Andrey M. Zankovich $^{1}$ and Mikhail A. Eremin $^{1}$}

$^{1}$ \quad Institute of Mathematics and Information Technology, Volgograd State University, Universitetskij Prospekt, 100, Volgograd 400062, Russia; korolev.vv@volsu.ru (V.V.K.); mabezborodov@rambler.ru (M.A.B.); zed81@list.ru (A.M.Z.); ereminmikhail@gmail.com (M.A.E.)

$^{*}$ \quad{Correspondence: ilya.g.kovalenko@gmail.com; Tel.: +7-8442-460812}

{The aim of this work is to construct an algorithm for visualizing a polydisperse phase of solid particles (dust) in an inhomogeneous flow of a two-phase gas-dust mixture that would allow us to see, within one plot, the degree of polydispersity of the dust phase and the difference in the spatial distributions of individual fractions of dust particles in the computational domain.
The developed technique allows us to reproduce concentrations from one to three fractions of dust particles in each cell in the computational domain. Each of the three fractions of dust particles is mapped to one of the main channels of the RGB palette. The intensity of the color shade is set to be proportional to the relative concentration of dust particles in this fraction. The final image for a polydisperse mixture is obtained by adding images in each of the three color channels. To visualize the degree of polydispersity, I propose depicting the spatial distribution of the entropy of the dust mixture. The definition of the entropy of a mixture is generalized to take into account the states of a mixture with zero number of particles in the mixture. They correspond to dust-free sections of the computational domain (voids). The proposed method for visualizing the polydispersity of a mixture of particles is demonstrated using the example of dynamic numerical modeling of the spatial features of dust structures formed in turbulent gas-dust flows and in flows with shock waves.}


\section{Introduction}

Real liquid, gas or plasma media have an inhomogeneous structure and are complex, composite systems that include microobjects of different types and different nature, such as small solid suspended particles, droplets or bubbles. There are a huge number of areas in the field of technology in which a person deals with multiphase environments, from chemical reactors to engine building and aeronautics. There are no manifestations of polyphase in nature that are less significant. Despite the small relative mass content, impurity aerosol or dust particles can play an important role in the life of planetary atmospheres, interstellar or intergalactic gas. The dust component determines the optical properties of the medium, its opacity in one interval or another of the electromagnetic spectrum. As a consequence, dust in the atmosphere or in the interstellar gas can act as a coolant, realizing the anti-greenhouse effect \cite{Krugel,Robock}. Due to their windage, dust grains can cause an effective mechanical effect of star radiation on a  transparent neutral gas, accelerating under the influence of radiation pressure and accelerating the surrounding matter \cite{BiaFerrara05,Zhukova}. Particles of dust in the interstellar gas act as a catalyst for the process of gas molecularization \cite{Krugel}.
Dust also serves as a building material for the formation of solid celestial bodies, such as asteroids or planets \cite{Krugel}.

Examples of space objects enriched with impure dust particles are gas-dust interstellar clouds and nebulae, protoplanetary disks, spiral arms of galaxies and gas-dust halos of galaxies. Usually, cosmic dust grains are particles with a size from millimeters to tens of nanometers \cite{Draine2004}. In the latter case, large molecules of the polyaromatic hydrocarbon (PAH) type are also classified as impurity nanometer particles \cite{Krugel}. On large scales, ranging from the sizes of stellar systems up to galactic ones, macroscopic objects such as small celestial bodies (meteoroids, asteroids, comet nuclei, planetesimals) can be understood as microscopic particles in a cosmic medium.

The dust component of interplanetary, interstellar and intergalactic media is a conglomerate of particles of various sizes and shapes, chemical composition, electric charge, optical and dielectric properties of the surface, etc. The behavior of various types of dust particles accelerated by the gas flow, by the radiation pressure force or by the Lorentz force, depends differently and substantially on the listed features of the structure of the particles. Thus, other things being equal, the effectiveness of sweeping efficiency of stars from the circumstellar space of graphite and silicate particles differs by 2--3 orders of magnitude \cite{V83}. The dependence on the shape of the particle (spherical or elongated aspherical) is weaker, but also leads to appreciable variations in the balaying efficiency values by approximately half the order of \cite{V83}. The presence of a particle size distribution can substantially distort the dispersion properties of waves in a cosmic plasma \cite{Shch_Prud} in comparison with a monodisperse gas-dust medium. A charged polydisperse dust component in the plasma leads to the formation of structures in the plasma (this phenomenon has been called Coulomb crystal) \cite{Fortov10,Merlino}. The dust charge determines the degree of its freezing into the surrounding gas, since the frictional force of the dust particle about the gas (Coulomb or collisional) can differ by orders of magnitude \cite{Draine79}. All this leads to the fact that the dust medium does not reproduce exactly motion of the carrier phase. Different dust fractions move relative to the gas in a heterogeneous manner, depending on the type of dust. Variation of motions of dust grains of different sorts lead to the formation of spatial variations of various dust fractions, which manifests itself as disproportions in the distribution of dust particles by varieties at different points of space.

Among the interesting phenomena from the physical point of view are dynamically conditioned processes (for example, during centrifugation in turbulent vortex cells). These can be classified as particle filtering, segregation by physical properties \cite{Marchisio13} or formation of areas free from dust (voiding) \cite{Fedoseev16}. 

An important feature of dust is that it can serve as a tracer for the morphological features of astrophysical objects. Thus, a thick dust lane in the spiral arm of a disk galaxy is usually associated with a galactic shock wave (GSW), because the dust lanes that extended along the arm and inside the arm coincide with the position of active formation of young stars possibly stimulated by the galactic shock front \cite{Roberts69}. Nevertheless, if we consider that gas and dust have different dynamic properties, then the statement that the dust lanes coincide with the position of the shock front is not obvious. However, not only the concentration of dust particles in individual areas of space, but also the difference in the spatial distributions of particles of different varieties is important for the diagnostics of the space environment. For example, the observed spatial variations of dust particles in size in galaxies \cite{Gon_arm, Gon_local, Gon_out, Zas} can serve as a theoretical tool for verification of some dynamic models of galaxies or interstellar medium.

One way to describe the structure and dynamics of multiphase media is to represent the medium as a mixture of the carrier of the continuous phase and the dispersed phase of the impurity particles. Impurity particles may have a spread in their properties. In this case, we talk about the polydispersity of the medium. In general, the polydispersity of impurity particles is understood as the difference in their physical and chemical characteristics (size, shape, mass, electrical charge, chemical composition) or motion parameters (speed). In the present paper we confine ourselves to the definition of polydispersity as a property of the inequality of particles in size.

Numerical modeling of gas-dust polydisperse medium flows establishes the connection between theory and observations. The results obtained in numerical simulation require clarity for analysis. Visual representation of data in graphic form greatly facilitates their comparison with experiment, observations and interpretation. The visual representation of multivariate data becomes particularly important in the modeling of processes in the interstellar medium. Information obtained from astronomical observations gives a limited idea of the structure of the objects under study, since light comes to the observer from various sources located at different distances and at the same time in one line of sight. Dynamic numerical modeling and visualization of its results allows us not only to restore the instantaneous three-dimensional structure of these objects, but also to consider their evolution.

An important issue is the possibility of combining in one figure images of the distributions of several fractions of particles, while retaining information on the distribution of dust particles within each fraction, since such visualization is convenient for comparative analysis. There is a method of approximate description of the degree of polydispersity of the medium through the use of a single parameter, the polydispersity index (PDI), defined as the ratio of the weighted squares of the sizes and the squared of the suspended particle sizes \cite{Palangetic14, Fortov10}
\begin{equation}\label{PDI}
  PDI = \frac{\left(\sum_i{N_i a_i^2}\right)\left(\sum_i{N_i}\right)}{\left(\sum_i{N_i a_i}\right)^2},
\end{equation}
where $N_i$ is the quantitative content of particles with sizes $a_i$. However, this reduced one-parameter description of polydispersity is probably too limited for a full analysis.

On the other hand, there are ingenious graphical tricks of multivariate data visualization \cite{Chernoff}, that are inconvenient for fast decoding of visual structures. An important requirement for the visualization procedure is the condition of perception of information at a glance. The usual method of perceiving graphic information is based on the use of the color structure of the image. Since the color palette is three-dimensional, we can use color as a resource for visualization of three-component data. In this paper we propose a method for simultaneous visualization of spatial distributions of several components of a polydisperse dust medium if the number of components does not exceed three. The idea of the method is based on our work \cite{KK2013}.

Section~\ref{polydisperse_dust} lists the properties of the dusty interstellar medium, the procedure for visualization of the polydisperse mixture is described in~\ref{visualization}, and the basic equations of motion of gas and dust are given in~\ref{HydroEqs}. In sections \ref{Dust_in_Clouds} and \ref{GUV} the application of the above methodology for the visualization of gas-dust flow in a turbulent cloud and in the vicinity of galactic shock waves is considered. The conclusion summarizes the main results.

\section{The Model of Polydisperse Dust Component of Gas-Dust Mixture}\label{polydisperse_dust}

A key characteristic that determines the distinction between dust particles by the grades in our model is the size of the dust particles.

As a rule, the dust-particle distributions observed in the interstellar medium have a maximum in the range $a_{min} \sim 0.001 \div 0.005$ $\mu$m \cite{MRN, WeinDr} and decrease with increasing radius. Large dust particles with radii $a \gtrsim a_{max}$ in a diffuse interstellar medium are rapidly destroyed, unless there is an intensive supply of their quantity by any sources. The size distribution function, therefore, has a sharp decline at $a>a_{max}$. The authors \cite{MRN} adopt the cutoff radius $a_{max}=0.25$ $\mu$m. Their followers  \cite{WeinDr} leave this value for silicate particles, and increase $a_{max}$ to $\sim 1 \div 10$ $\mu$m for carbon particles.

The dependence of the concentration of particles per unit interval of their radii $a$ in the range $a_{min}<a<a_{max}$, which decreases monotonically with increasing particle sizes, is often approximated by the power function $n_d (a) \sim a^{-r}$. The authors of the pioneering work \cite{MRN} assume $r=3.5$. Subsequent work \cite{Draine2004, Hira_etal2014} refines this approximation, offering more complex dependencies, other than power.

In the regions of increased gas density (molecular cloud nuclei, condensation in a turbulent gas) with gas concentrations $n_g\gtrsim 10^5$ cm $^{-3}$, the effects of coagulation of dust particles and the accretion of molecules on the surface of dust particles lead to the intense formation of large dust grains of micron size
\cite{Hira_Yan2009}.

For our purposes, the use of very detailed dust particle distribution sizes is excessive.
We consider either the dynamics of each fraction separately, or an equilibrium dust mixture.
The polydispersity of the dust component is simulated by an ensemble of particles consisting of three fractions, in each of which the particle sizes and masses are fixed and identical.

The models of the gas-dust interstellar medium, which we are considering in this paper, are related to the presence of dense clouds. In one case, we model directly the inner regions of turbulent molecular clouds. In the other, we consider the flow of a gas-dust interstellar medium through spiral arms, to  which, as is well known, there is a maximum concentration of molecular clouds in galaxies.
This gives us reason to consider the motion of not only the most common dust grains with dimensions of the order of $0.1$ $\mu$m, but also large micron dust particles. Quite small particles with dimensions $a \sim 0.01$ $\mu$m have too small a length of dynamic relaxation in our problem, in which characteristic spatial scales are 1 pc in one case and 1 kpc in the other, so the motion of such particles almost completely replicates the motion of the gas and we do not include them in the consideration.

Dynamic relaxation is understood as the process of equalizing the velocities of dust particles and gas due to the action of frictional force. The degree of dynamic coupling of dust and gas is characterized by the dimensionless Stokes number, defined as the ratio of the dynamic relaxation time $\tau_{fr}$ to the characteristic dynamic time $\tau$ of the problem:
\begin{equation}\label{Stokes_def}
{\rm Sk}= \frac{\tau_{fr}}{\tau}.
\end{equation}

From the dynamic point of view, the most interesting properties are those with Stokes numbers ${\rm Sk} \sim 0.1 \div 10$. In this case, the degree of clustering of the dust component in the turbulized gas is maximal. In the calculations, we consider dust particles with Stokes numbers equal to $0.05$, $0.5$, and $5$, which corresponds to the radii of dust particles $a=0.15$, $1.5$ and $15$ $\mu$m, respectively. Particles whose radius is $a=15$ $\mu$m are conventionally called large, particles with radii $a=1.5$ $\mu$m are considered to be medium-sized dust particles, and the smallest particles in calculations with radii $a=0.15$ $\mu$m are small.

To simplify the description, we do not distinguish dust particles by chemical composition, we consider them as solid graphite beads with a density of 2.23 g/cm$^3$ \cite{Krugel}.

\section{Visualization Procedure of Spatial Distribution of Polydisperse Dust}\label{visualization}

The choice of three dust particles in the model under consideration is due to the possibilities of using a three-dimensional color RGB palette to visualize the spatial distribution of dust particles of various sizes. Visualization is of interest for solving the problem of identifying the physical mechanisms responsible for the segregation of dust particles in size, which is actually observed in the interstellar medium.

The visualization procedure is focused on the perception of a color image by the human eye.
The real spectrum of electromagnetic radiation in the optical range is continual and any color in the general case is an infinite uncountable sum of monochromatic components of the spectrum. The peculiarity of the structure of the human visual system is that the overwhelming majority of colors or hues are capable of perceiving it as a mixture of a small number of primary colors. Many years of practice shows that it is enough to use the three main color channels. In the RGB palette, these are red, green and blue.

The decomposition of an arbitrary color over three basic colors mathematically corresponds to the spectral or, in the terminology of quantum mechanics, the ket-bra-decomposition of an arbitrary three-dimensional matrix on three mutually orthogonal basis three-dimensional rank-one matrices with real positive semi-definite coefficients of decomposition.
This decomposition is analogous to the spectral representation of the density matrix in quantum mechanics or in statistical physics. Any color or shade in the RGB palette can thus be matched with some matrix whose rank is one if the color is pure or more than one if the color is composite. The maximum rank of a three-dimensional matrix is three, in which case the mixture is composed of all three basic colors.

In the literature, the color state is sometimes not attributed to the RGB matrix, but the RGB vector, and the expansion, respectively, are constructed as a basis vector decomposition. The definition of a color state through vectors is not very successful, because it implicitly assumes superposition of colors in view of their interference, while the eye does not perceive interference. The addition of colors should be considered as an incoherent mixture and described in the superposition language of the matrices.

In accordance with the terminology adopted in quantum mechanics, the amplitudes are called the coefficients of the expansion of vectors, so the term `amplitude' brings to mind coherent superposition. To avoid misunderstandings for describing the spectral decomposition of matrices, we use the term `decomposition coefficient' instead of the term `amplitude'.

The relationship between the coefficients of the spectral decomposition of matrices determines the color, and their absolute values --- luminance. Any set of three real non-negative coefficients can thus be characterized by a certain color and brightness.

Basic colors are usually called chromatic, and any other color, obtained by a mixture of basic colors, is partially or completely achromatic. The maximum achromatic colors corresponding to the balanced mixture of colors are shades of gray.

If we fix the ratios between all three coefficients of the RGB decomposition, and then simultaneously increase or decrease them the same number of times, we get a certain scale containing different levels of brightness of the same color. This scale can be compared with a one-dimensional color palette, which is commonly called the gradation of brightness. An example of this kind of one-dimensional color palette is the shades of gray. With a mixture of chromatic color and achromatic, the corresponding one-dimensional palette can express the gradation of saturation. Example: a transition from pure color to white.

Chromatic colors are associated with pure states of the mixture, and achromatic colors are mixed. Thus, for distributions of three-component dust, a clean state in a certain spatial cell of the computational domain corresponds to a situation when in this cell there are dust particles of one and only one fraction. Otherwise, the state of the dusty medium in this cell is called mixed.

The measure of the degree of mixed state is the entropy of the state.
Entropy is minimal if the state is pure, and maximal when the state is a homogeneous mixture.

\subsection{Gas-Dust Mixture Visualization}\label{visual_mix}

The idea of color visualization of a mixture characterized by three arbitrary real weight coefficients was realized in the paper \cite{KK2013} for determining the position and identification of the types of hydrodynamic jumps (shock waves or tangential discontinuities), as well as rarefaction waves in arbitrary flows of a continuous medium.

The authors of \cite{KK2013} proposed each type of hydrodynamic discontinuity (tangential, shock) or rarefaction waves to match a certain color, and the intensity of the color is the amplitude of the shock or the rarefaction wave. The type and intensity of the shock or rarefaction waves characterize three eigenvalues of the strain-rate tensor of the medium that make up the complete system of its functionally independent invariants.

Similarly, one can proceed in the case of visualization of spatial distributions of dust, using the available degrees of freedom `color-intensity'.

Each of the three fractions of dust particles is associated with one of the main channels of pure color: red, green or blue. The intensity of the pure color $c_i (x, y)$ in the color channel in the $i$ color space in a certain cell will be determined from the dust concentration in the same cell. In calculations, under the dust concentration $n(x, y)$, we mean the number of particles in a cell. The color tone is a function of the color coordinates whose values lie in the unit interval $ [0,1]$.
One way to calculate the color tone is to normalize the local value of the concentration of the $i$-th fraction $n_{i}(x, y)$ to the global maximum value of the concentration of this fraction $\max(n_{i})$, determined by all cells of the computation area:
\begin{equation}\label{c_i}
c_i(x,y) = \frac {n_{i}(x,y)} {\max(n_{i})},  \quad i=\overline{1..3}.
\end{equation}

If you want to detail the features of the distributions, you can use various contrasting transformations, for example, to amplify the low-amplitude component of the signal, a hyperbolic tangent
\begin{equation}\label{c_i_tanh}
\tilde{c}_i(x,y) =  \frac {\tanh( K n_{i}(x,y) )} {\tanh( K \max(n_{i}) )},  \quad i=\overline{1..3}
\end{equation}
with a scaling parameter $1\lesssim K \lesssim 10$. An even simpler way to remove the low-amplitude component is the threshold image filtering used in the construction of the Figures \ref{Fig2}, \ref{Fig4}.

\begin{figure}[ht]
\begin{minipage}[ht]{1.0\linewidth}
\centering{\includegraphics[width=1.0\linewidth]{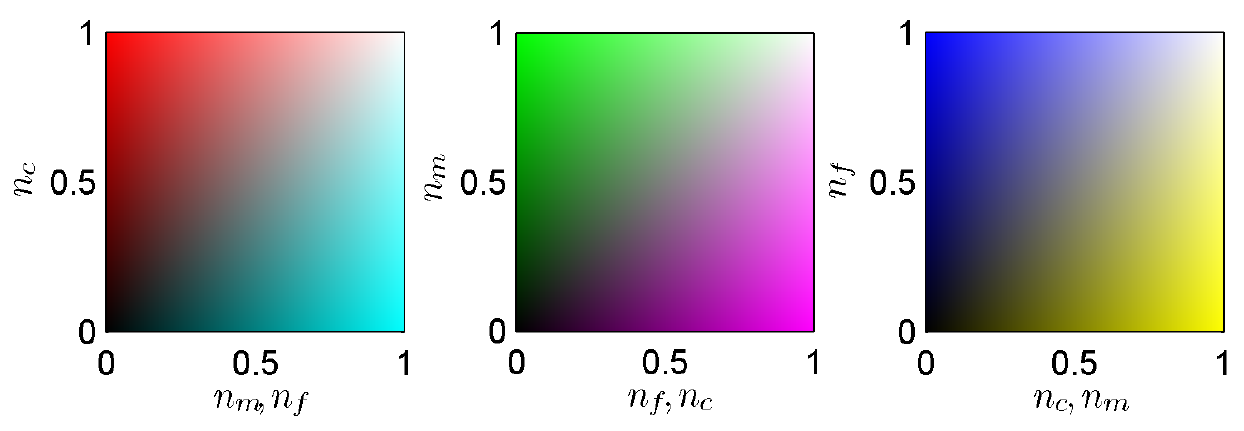}}\\
\caption{
Palettes of colors correspond to different combinations of relative concentrations of three dust fractions. The subscripts $c$, $m$, $f$ correspond to the concentrations of fractions of coarse, medium-sized and fine grains, respectively. The sections of the cubic color palette are shown, which correspond to the condition that the two fractions (their concentration values are marked on the abscissa axis) have the same concentration. The area in the upper left corner corresponds to the predominance of one fraction of dust particles over the other two. The right lower corner corresponds to the predominance of two dust particles simultaneously over the third one. Colors in the lower left corner correspond to a low concentration of all three fractions. The colors in the upper right corner correspond to the concentrations of all three fractions, close to the maximum. The state of the mixture, in which the concentrations of all three fractions are the same, are represented by shades of gray---from black, when the concentrations are zero, to white, when the concentrations are maximum.
}
\label{Fig1}
\end{minipage}
\end{figure}

\begin{figure}[h]
\begin{minipage}[h]{0.485\linewidth}
\center{\includegraphics[width=1\linewidth]{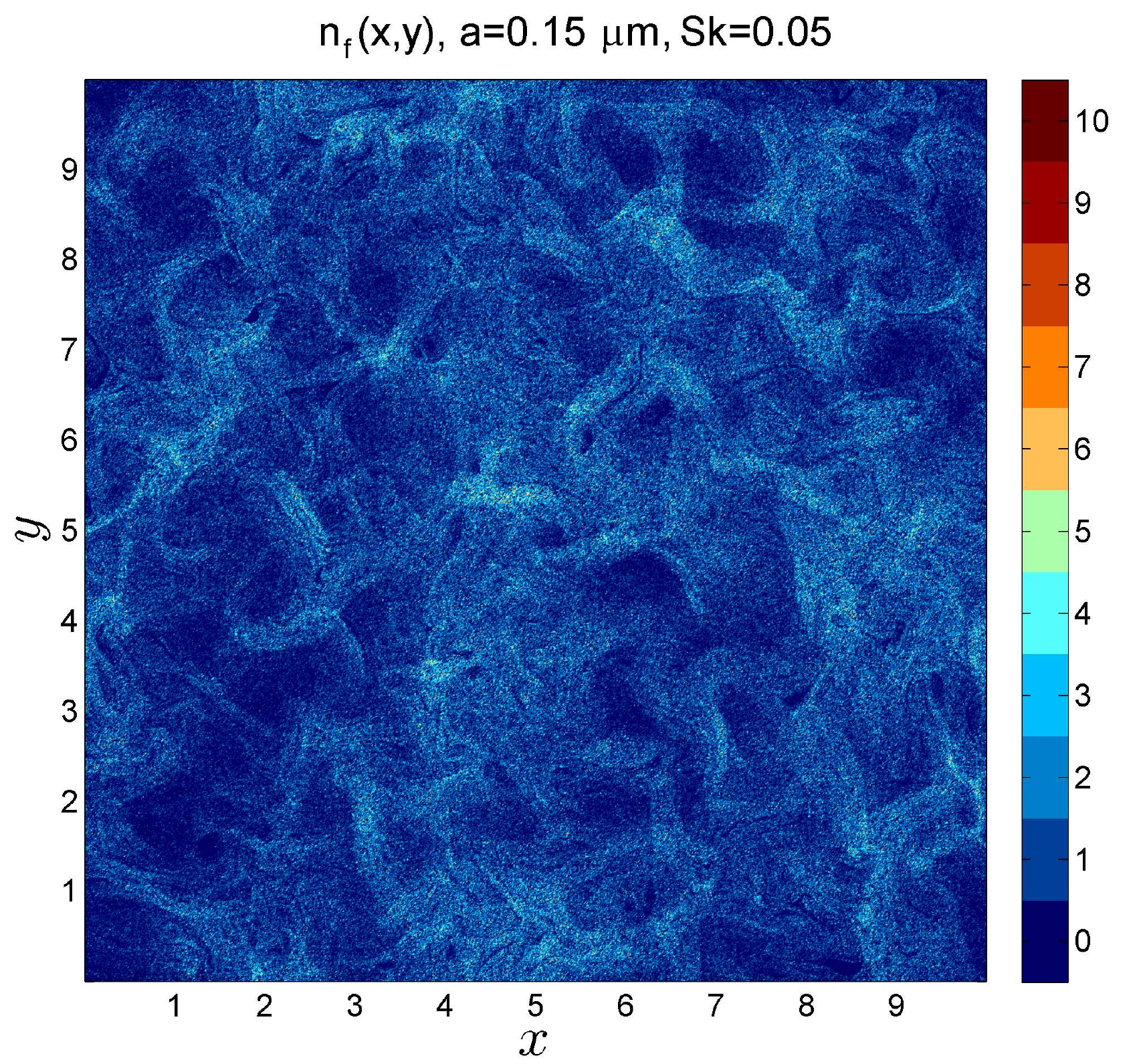}} a) \\
\end{minipage}
\hfill
\begin{minipage}[h]{0.485\linewidth}
\center{\includegraphics[width=1\linewidth]{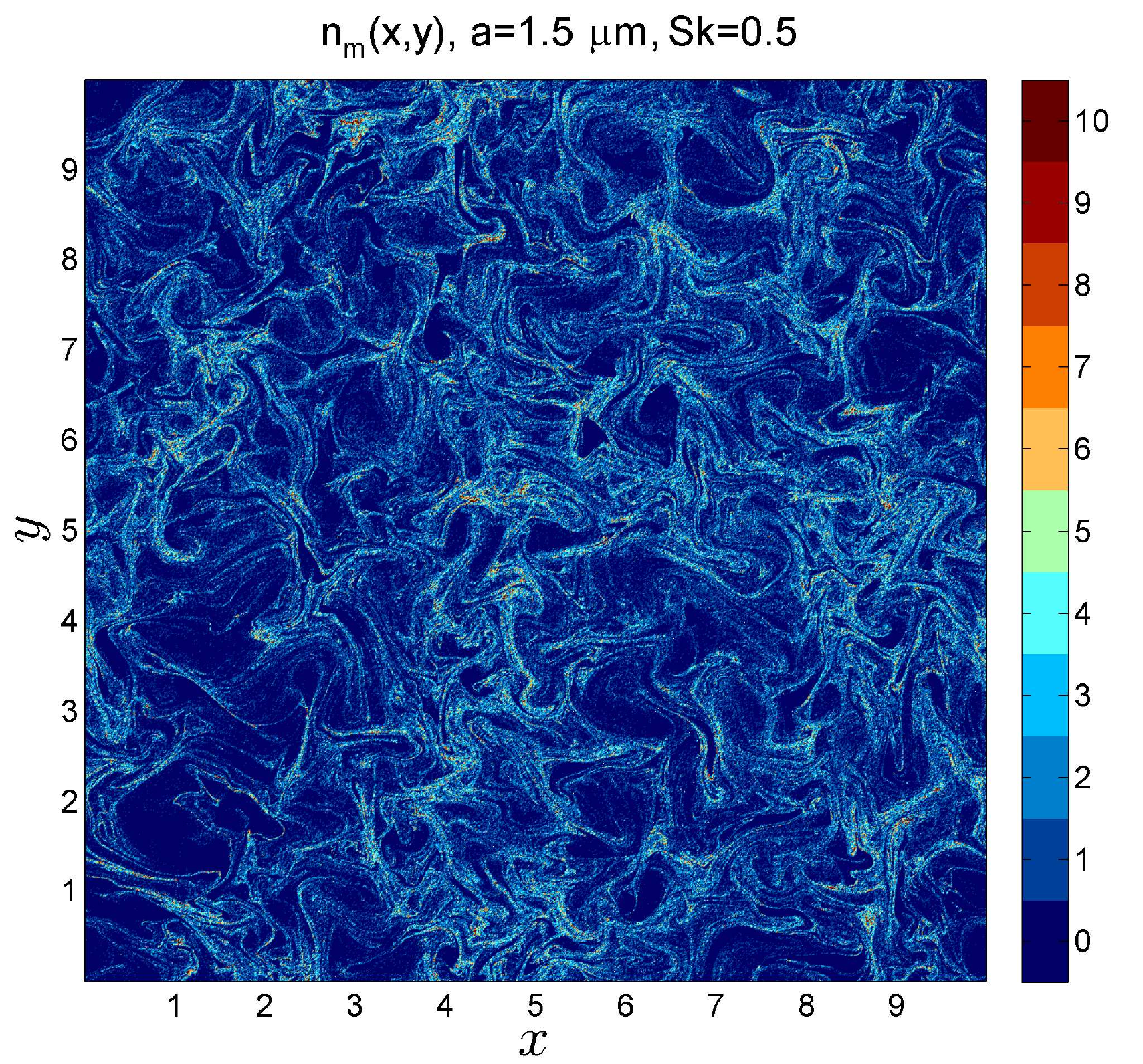}} \\b)
\end{minipage}
\vfill
\begin{minipage}[h]{0.485\linewidth}
\center{\includegraphics[width=1\linewidth]{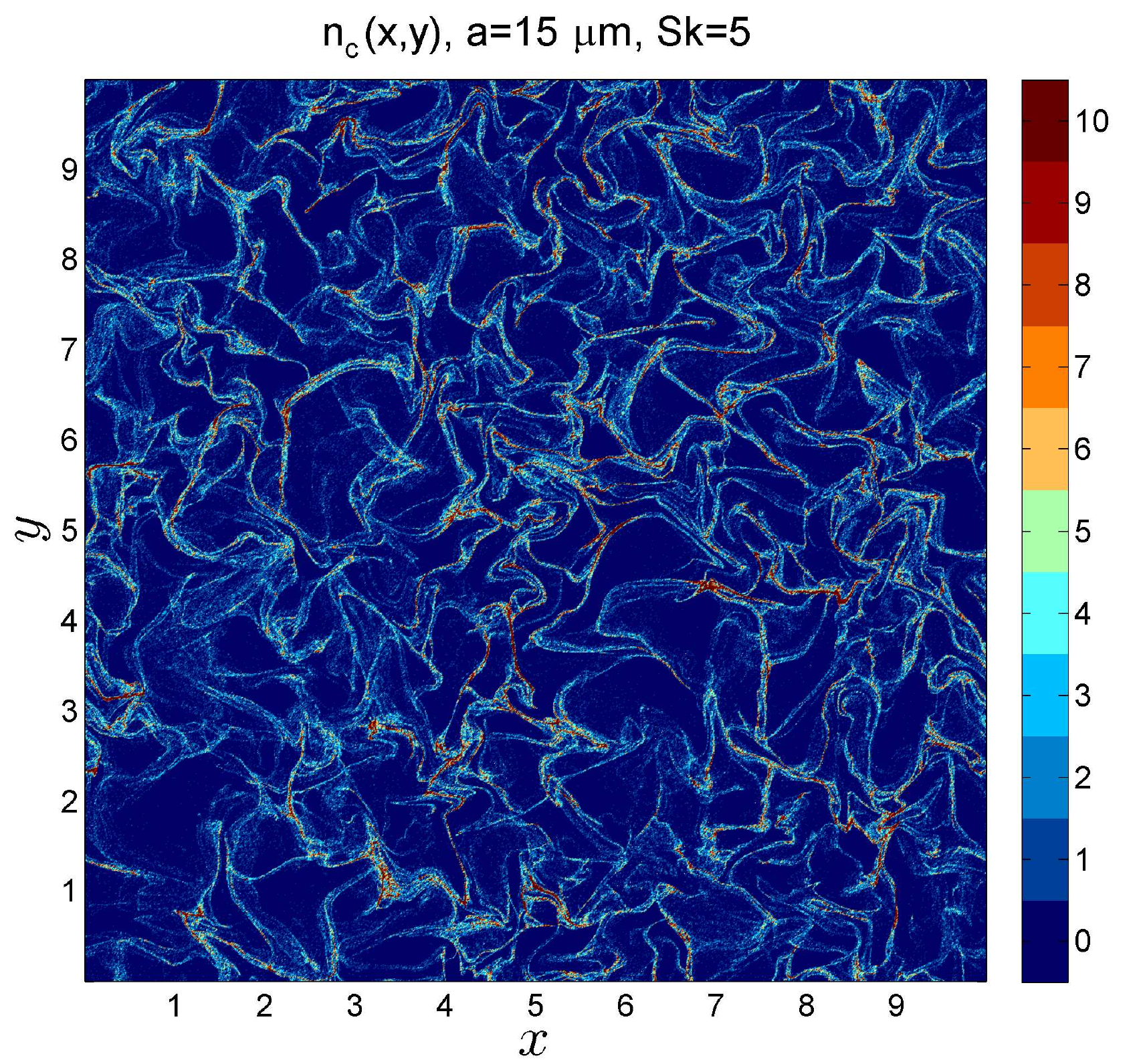}} c) \\
\end{minipage}
\hfill
\begin{minipage}[h]{0.485\linewidth}
\center{\includegraphics[width=1\linewidth]{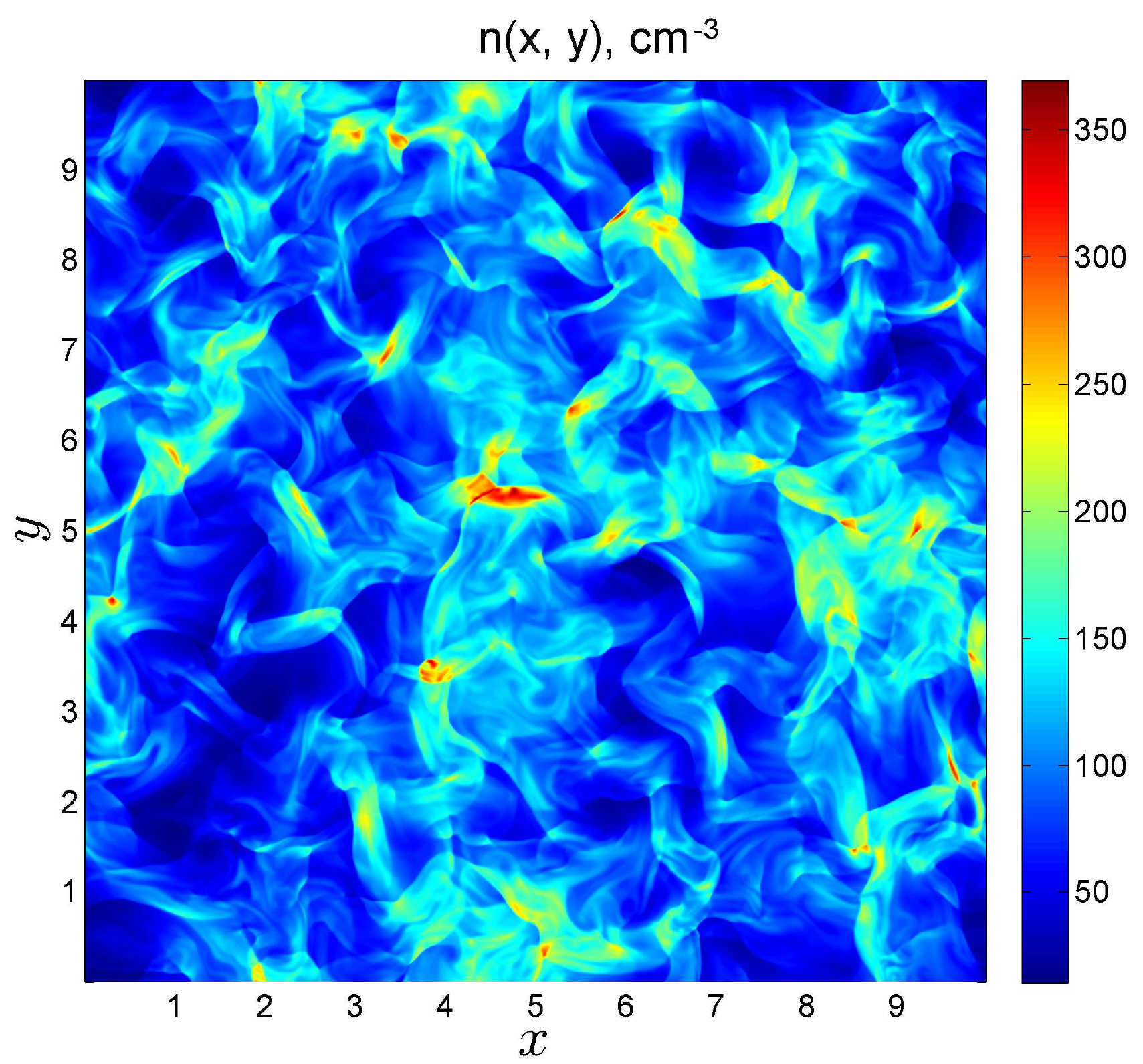}} d) \\
\end{minipage}
\caption{
The distribution of the concentrations of individual dust fractions at the same time in a turbulent molecular cloud model for particles of different sizes is: (a) $a=0.15$ $\mu$m, (b) $a=1.5$ $\mu$m, (c) $a=15$ $\mu$m. The corresponding Stokes numbers $Sk$ are given in the headings to the figures. The discrete color scale on the vertical bar compares the color to the number of particles in the cell.
The distributions are cut off from above by the maximum displayed number of particles: if the number of particles in the cell is greater than ten, then the number of particles in the figure is displayed as ten. Figure (d) shows the distribution of gas concentration at the same instant.
}
\label{Fig2}
\end{figure}

The resulting image is obtained by adding images in each of the three color channels.
For example, if at a given point the relative amount of dust of all types is the same, then a combination of $[c_1, c_2, c_3]$ will give a certain shade of gray at this point.
The resulting color will also visually be a shade of gray, if the values of $c_i$ are close in value to each other. Black color will be obtained under the condition $c_1 = c_2 = c_3 = 0$ in cells where a global minimum of the relative dust content is achieved for all three fractions.
The white color corresponds to the global maximum of the relative content, i.e., $c_1 = c_2 = c_3 = 1$. If the content of only one kind of dust is predominant relative to the other two, the point is colored in one of the three main chromatic colors corresponding to this sort of dust.
Conversely, with a small amount of dust of one of the varieties, the color of the cell will be determined by the other two as an additional color; i.e., it will be a shade of yellow, cyan or magenta (see Figure~\ref{Fig1}).

The proposed technique for visualizing the spatial distribution of polydisperse dust is hereinafter referred to as the tricolor technique.

\subsection{Visualization of the Gas-Dust Mixture Entropy}\label{visual_entro}
The entropy of the mixture is characterized not by three numbers, but by one number, so to visualize it, it is sufficient to use a one-dimensional palette, for example, grayscale.

The entropy of a system that can be in different states is interpreted as a measure of randomness in measuring the state of the system. With respect to a mixture of particles of three sorts, the states found in a certain cell $(x, y)$ as a result of the measurement correspond to three possible realizations: ``the particle of the first grade'' is found, ``the particle of the second grade'' is found, ``the particle of the third grade'' is found. The entropy of a mixture of particles is usually defined in terms of the probabilities $p_i (x, y)$ for detecting particles of the $i$-th kind. In our case, it is natural to specify the entropy as measured in trits
\begin{equation}\label{entropy_def_3}
  S(x,y) = - \sum_{i=1}^3 p_i(x,y) \log_3{p_i(x,y)}.
\end{equation}
The maximum of entropy occurs in the case of an equilibrium mixture, this corresponds to one trit, the minimum corresponds to a pure state when only one species of particles is present in the mixture. In this latter case, the result of the measurement is predictable with absolute certainty and the entropy is zero.

Specificity of the standard definition of the entropy of the mixture \eqref{entropy_def_3} is that it assumes the fulfillment of the law of conservation of total probability: the probability of detecting at least some particle of the mixture is unity,
\begin{equation}\label{sum_p_i_1N_stand}
 \sum_{i=1}^3  p_i(x,y) = 1.
\end{equation}
In our problem, the situation is more complicated, since one more possible state can be realized, when there is not a single particle in the cell at all.
In this case, obviously, one should assume
\begin{equation}\label{sum_p_i_13=0}
p_1(x,y) = p_2(x,y) = p_3(x,y) = 0,
\end{equation}
and then the completeness condition for the set of realizations of \eqref{sum_p_i_1N_stand} is not fulfilled. Accordingly, the expression for entropy \eqref{entropy_def_3} ceases to be fair. Formally, when the condition \eqref{sum_p_i_13=0} is satisfied, the entropy \eqref{entropy_def_3} is equal to zero, but then two qualitatively different cases of a mixture of empty and a monodisperse mixture are identified, which is obviously unacceptable for visualization purposes.

The need to distinguish between an empty and a monodisperse mixture forces us to introduce an extended description of the state of the mixture. We supplement the number of possible realizations when measuring the type of a particle in a cell by the fourth possible result, ``no particles found''. The probability of such a measurement result is denoted as $p_0$. Our proposal is to consider the absence of particles in a cell as the presence of a quasiparticle, that is, a hole. In this description, the mixture is considered as consisting of four fractions, three of which fall on real particles, and the fourth one is hole fractions.

The normalization condition for the total probability with allowance for the holes is now restored,
\begin{equation}\label{sum_p_i_0N}
 \sum_{i=0}^3  p_i(x,y)
= 1.
\end{equation}
The probability $ p_0 $ of the particle's non-detection is either equal to zero if there is at least one real particle in the cell, and, therefore, there are no holes in the cell, or one if there is a hole in the cell. The partial sum of three probabilities $p_i$, $i=\overline {1..3}$, now determines the probability of finding at least one particle in the cell.

The probabilities can be calculated in terms of the relative concentrations $n_i$ of the content of $i$-th particles in the cell. The concentration of holes is determined as a unit if there are no particles in the cell, and zero if there is at least one particle present,
\begin{equation}\label{n_0}
 n_0(x,y) = 1- {\rm sgn}\left(\sum_{i=1}^3  n_i(x,y)\right).
\end{equation}
The function ${\rm sgn}(z)$ used here is zero at zero.

The probabilities of detecting a $i$-th particle or not detecting a single particle are determined through their concentrations, taking into account \eqref{n_0} by the standard formula,
\begin{equation}\label{p_i}
  p_j(x,y) = \frac{n_j(x,y)}{\sum\limits_{i=0}^3 n_i(x,y)}, \quad j=\overline{0..3}.
\end{equation}
Such a definition of probabilities guarantees their real definiteness, belonging to the unit interval $[0,1]$, and the completeness of \eqref{sum_p_i_0N}.

The mixture can thus be defined and regarded as a composite system of the $PH$ particle-hole system in which the states of the $P$ (``particle'') and $H$ (``hole'') are completely anticorrelated. If a hole is found in the cell, then there are no particles in this cell, and vice versa.

We introduce the conditional probabilities $p_i(P|H=h_j)$, $i=\overline{0..3} $, $j=\overline{0..1}$, as the probability that the $i$-th particle is not detected either, provided that the hole in this cell is detected in the state $h_j$. By $h_0 \equiv 0$ we mean a state where there is no hole, under $h_1\equiv 1$, when there is a hole. In accordance with \eqref{n_0} and \eqref{p_i}, depending on whether there is a hole in the cell or not, we have two different conditional probability distributions. The corresponding distributions are given in Table~1.
The first distribution corresponds to a non-empty mixture, the second is empty.

\begin{table}[h]\label{table1}
\caption{Conditional probability $p_i(P|H=h_j)$ of detection or non-detection of particles in cell $(x, y)$ and conditional entropy $S_j$ for various outcomes of hole detection.
}
\centering
\begin{tabular}{cccccc}
\toprule
        & $p_0(P|H=h_j)$ & $p_1(P|H=h_j)$ & $p_2(P|H=h_j)$ & $p_3(P|H=h_j)$ & $S_j$ \\
\toprule
  $h_0$ & 0 & $p_1(x,y)$ & $p_2(x,y)$ & $p_3(x,y)$ & $S_0=S$ \textrm{(Eq.~\eqref{entropy_def_3})} \\
  $h_1$ & 1 & 0          & 0          & 0          & 0           \\
\toprule
\end{tabular}
\end{table}

To distinguish the cases of empty and monodisperse mixtures when visualizing the spatial distribution of the entropy of the mixture, we propose, instead of the total entropy expressed by one number, to consider the set of conditional entropies given as their distribution over various realizations of the state of holes. These conditional entropies determine the entropy of the subsystem $P$ for a particular measurement result of the subsystem $H$.
If the cell is filled with particles and holes are not present, $p_0=0$, then the conditional entropy corresponding to this implementation is $S_0$, if there is a hole, $p_0=1$, conditional entropy takes the value $S_1$. Both conditional entropies $S_0$ and $S_1$ are calculated by a single formula
\begin{equation}\label{entropy_def_4}
  S_j(x,y) = - \sum_{i=0}^3 p_i(P|H=h_j;\,x,y) \log_3{p_i(P|H=h_j;\,x,y)}, \quad j=\overline{0..1}.
\end{equation}

We match our one-dimensional color palette to each of the two conditional entropies.
In the event that there is at least one particle in the cell, we set the palette as a grayscale palette. The conditional entropy $S_0$ in this case is exactly equal to the entropy calculated by the formula \eqref{entropy_def_3} and taking values in the interval from zero to one. For better perception of the image, we invert the color scale. To the value $S_0=1$ we assign not white, but black color.

For cells filled with holes, we attribute an alternative one-dimensional palette as a gradation of the saturation of blue, namely, a mixture of blue and white in one direction or another. For the convenience of visualization, this palette is also inverted: the minimum of entropy is the maximum saturation.
From the definition of \eqref{entropy_def_4} and the data from Table~1 it follows that in the presence of a hole in the cell, the state of the subsystem $P$ is always pure, the value of the conditional entropy $S_1$, therefore, is always the same and equal to zero. Accordingly, in practice, in drawings, instead of the potentially possible gradations of blue, only one color is realized, the brightest blue as possible.

\section{Equations of Gas-Dust Mixture Dynamics}\label{HydroEqs}
The motion of the interstellar gas as a collisional continuum is described by a system of equations
\begin{equation} \label{hydro_1}
    \frac{\partial \rho}{\partial t}+ \nabla(\rho \textbf{v})= 0,
\end{equation}
\begin{equation} \label{hydro_2}
    \frac{\partial}{\partial t} (\rho \textbf{v}) + \nabla(\rho \textbf{v} \otimes \textbf{v}) = -\nabla p - \rho\nabla\Phi,
\end{equation}
\begin{equation} \label{hydro_3}
    \frac{\partial E}{\partial t} + \nabla\left((E+p)\textbf{v}\right)= 0,
\end{equation}
\begin{equation} \label{hydro_4}
     E =\frac{\rho v^2}{2} + \frac{p}{\gamma-1} + \rho \Phi \;,
\end{equation}
\begin{equation} \label{hydro_5}
     p={n k_B T },
\end{equation}
where $\rho$ and $n$ are the gas density and concentration, $p$ is the pressure, $\mathbf{v}$ is the velocity, $E$ is the total energy of the unit volume of gas, $T$, gas temperature,
$\Phi = \Phi(\mathbf{r})$ is the stationary external gravitational potential,
$k_B$ is the Boltzmann constant.
The intrinsic gravity of the gas is neglected in comparison with the gravitational field, which is more massive than the gas by the stellar component of the galaxy.

The gas is assumed to be adiabatic with the adiabatic exponent $\gamma$. In all models, a gas with $\gamma = 5/3$ was considered, which corresponds to either a monatomic gas or a polyatomic gas at temperatures below the degeneracy temperature of the rotational and vibrational degrees of freedom of the molecules.

Since the dust component in galaxies is less massive than gas (the characteristic value of the dust mass is $\sim 1$\% of the mass of the gas), the dynamic effect of dust on the gas is neglected.

The dust is considered as collisionless (with respect to collisions of dust particles with each other) medium. The basic forces that determine the motion of dust particles in the model under consideration are the friction force $\mathbf{f}_{fr}$ of a dust particle surrounding the gas, the gravitational force $\mathbf{f}_{g}$, and the radiation pressure force $\mathbf{f}_{rad}$. Dust grains are considered uncharged, so the friction of dust particles on the gas is considered as collisional kinetic. The contribution of the Coulomb interaction due to friction is absent and the interaction of dust particles with the magnetic field is also not taken into account. Then for a dust particle we can write the equations of motion in the form
\begin{equation}\label{dust_r_eq}
    \frac{d\mathbf{r}_d}{dt}=\mathbf{v}_d,
\end{equation}
\begin{equation}\label{dust_v_eq}
    \frac{d \mathbf{v}_d}{dt}=\mathbf{f}_d.
\end{equation}
Here, $\mathbf{r}_d$ and $\mathbf{v}_d$ are the radius-vector and velocity vector of the particle, $\mathbf{f}_d $ is the total specific force per unit mass, acting on the particle:
\begin{equation}\label{force_total}
\mathbf{f}_{d}= \mathbf{f}_{g} + \mathbf{f}_{rad} + \mathbf{f}_{fr}.
\end{equation}
The frictional force can be written in the form \cite{Draine2004}
\begin{equation}\label{force fr}
\mathbf{f}_{fr}= \frac{n a^2}{m_d} \Delta \mathbf{v} \sqrt{\frac{128\pi}{9}k_B T m_g+\pi^2m_g(\Delta \mathbf{v})^2},
\end{equation}
where $m_g$ is the mass of the gas particle, $\Delta \mathbf{v} = \mathbf{v} - \mathbf{v}_d$ is the gas velocity relative to the particle. Note that the dust particles of different sizes have different dynamic properties: larger particles have more inertia and have a greater time of equalization of velocities with gas. Examples of flow areas in which dust particles are separated in terms of velocities are zones of sharp changes in the gas density or its velocity, for example, shock fronts.
Also, a sharp difference in the motion of particles of different sizes can be observed in the field of action of radiation pressure forces. This difference is caused by different windages of particles of different sizes.

The characteristic time of dynamic relaxation of a dust grain due to friction is defined as the inverse coefficient of friction in the limit of small relative velocities of grains,
$\tau_{fr} = \underset{\Delta v \to 0}{\lim}(\Delta v / f_{fr})$.
The dynamic time $\tau$ in the problem is defined as the ratio of the characteristic spatial scale of the problem $L$ to the characteristic velocity of sound in the gas $c_s = (\gamma k_B T/m_g)^{1/2}$. Taking into account the definition of friction force \eqref{force fr} and Stokes number \eqref {Stokes_def}, we have
\begin{equation}\label{Stokes_number}
{\rm Sk}= \frac{3\sqrt{\gamma} m_d }{8\sqrt{2\pi } a^2 n L m_g} .
\end{equation}
Assuming the mean value of the density of the dust particles material \cite{Krugel} $\rho_d \approx 2.2$ g $\cdot$ cm$^{-3}$ and assuming the particles are spherical, we rewrite \eqref{Stokes_number} as
\begin{equation}\label{Stokes_number_estimate}
{\rm Sk} \approx 0.345\left(\frac{a}{ 1 \mu{\rm m} }\right) \left(\frac{0.1 {\rm cm}^{-3}}{n}\right) \left( \frac{1 {\rm kpc}}{L}\right).
\end{equation}

The Stokes numbers $Sk = 0.05$, $0.5$, $5$ correspond to dusts with dimensions $a \approx 0.15$, $1.5$, $15$ $\mu$m in the problem of a galactic shock wave with a characteristic arm scale $L = 1$ kpc and gas concentration $n = 0.1$ cm$^{-3 }$

For dust particles of the same dimensions moving in a turbulent atomic-molecular cloud, with the characteristic scale of the cloud $L = 1$ pc and the gas concentration $n = 100$ cm$^{-3}$, we have the same set of Stokes numbers.
The method of specifying the radiation pressure force $\mathbf{f}_{rad}$ will be described in the section \ref{GUV}.

\section{Spatial Variations of Dust in a Turbulent Gas-Dust Cloud}\label{Dust_in_Clouds}

Since dust is a catalyst in the production of molecular hydrogen in the interstellar medium,
it is of interest to analyze the spatial correlations of the distributions of dust and molecular hydrogen in the clouds. The initial data for the detection of the molecule $H_2$, as a rule, are estimates of the radiation intensity in the lines of the $CO$ molecules, since molecular hydrogen does not have lines that could be excited in a cold gas. The main problem in estimating the mass of a molecular gas is due to the fact that the molecular clouds in the $CO$ line are not very transparent, and radiation comes mainly from the surface layers of the cloud. In order to estimate the conversion factor, it is necessary to have independent methods for estimating the mass of the molecular gas. If we assume that the correlation between the spatial distributions of dust and molecular hydrogen is large, then the dust can act as a tracer for the regions of localization of molecular hydrogen $H_2$ in the clouds.
Thus, it is necessary to construct a dynamic numerical model of a gas-dust molecular cloud, which takes into account the formation of molecules $CO$ and molecular hydrogen $H_2$, to study the spatial variations of molecules. Since the presence of molecular hydrogen in a given region in the cloud is associated with the presence of dust, it makes sense to analyze the spatial variations of dust and molecules, determine the statistical regularities of occurrence, dynamic evolution and instantaneous physical and geometric characteristics of molecular condensations, and study correlations between the content of molecules ($H_2$, $CO$) and dust in molecular clouds on the basis of dynamic numerical simulation.

For a primary analysis of the processes occurring in giant molecular clouds, a model of a turbulent gas-dust medium was constructed without taking into account chemical evolution and thermal processes.
Calculations were carried out for a fragment of the medium with a size of $10 \times 10$ pc.
The gas in the cloud was considered as a neutral atomic, consisting only of hydrogen atoms.
The unperturbed gas concentration was set to $n = 100$ cm $^{-3}$, and the unperturbed gas temperature was $T \sim 100$ K. For such values of temperature and concentration, the Jeans scale exceeded 10 pc, so gravity was not taken into account in the problem.

The motion of the dust particles was calculated by integrating the equations \eqref{dust_r_eq}, \eqref{dust_v_eq}. The system of equations of gas dynamics \eqref{hydro_1} - \eqref{hydro_5} was solved using parallel code based on a conservative numerical scheme of the MUSCL TVD \cite{ToroEF} type of second-order accuracy in time and the third in space on smooth flow sections.

Turbulence in a gas was generated by means of a quasi-random perturbation of the velocity field. Pulsations and vortices in interstellar clouds could be traced by observations on scales from the size of the cloud to $\sim 0.1$ pc \cite{Heyer2004, Falgarone1992}. The nature of the mechanisms that support chaotic variations in speed is not fully understood. In principle, they can be caused by a number of physical mechanisms: self-gravitation, magnetic fields, the development of instabilities, the interaction of gas with radiation, etc. In practice, in particular, in numerical simulation, it is difficult to directly take into account their joint influence. However, there are heuristic methods, known as `forcing', allowing the formation of a turbulent flow with the required parameters. In our work we used the approach \cite{Dubinski1995, MLow1999}, according to which the turbulent velocity field is calculated from its spectral characteristics. In the perturbation spectrum, both incompressible and compressible turbulence components were taken into account together in the spirit of the work \cite{Federrath2010}. Taking into account the compressible component of turbulence is important, since with the dispersion of the gas velocities in molecular clouds to $\sim 10$ km/s, the sound velocities are $\sim 0.1-0.3$ km/s \cite{ElmegreenScalo2004-1}.

The results of numerical simulation show that dust under the influence of turbulent gas motion is distributed significantly non-uniformly, with particles of different varieties having qualitatively different spatial distributions depending on the Stokes number (Figure~\ref{Fig2}).
Particles having a smaller size and, as a consequence, possessing a small inertia, are distributed more evenly over the computational area. Large particles tend to concentrate in dense clumps along the periphery of the vortices.

\begin{figure}[h]
\begin{minipage}[ht]{1.0\linewidth}
\center{\includegraphics[width=1.0\linewidth]{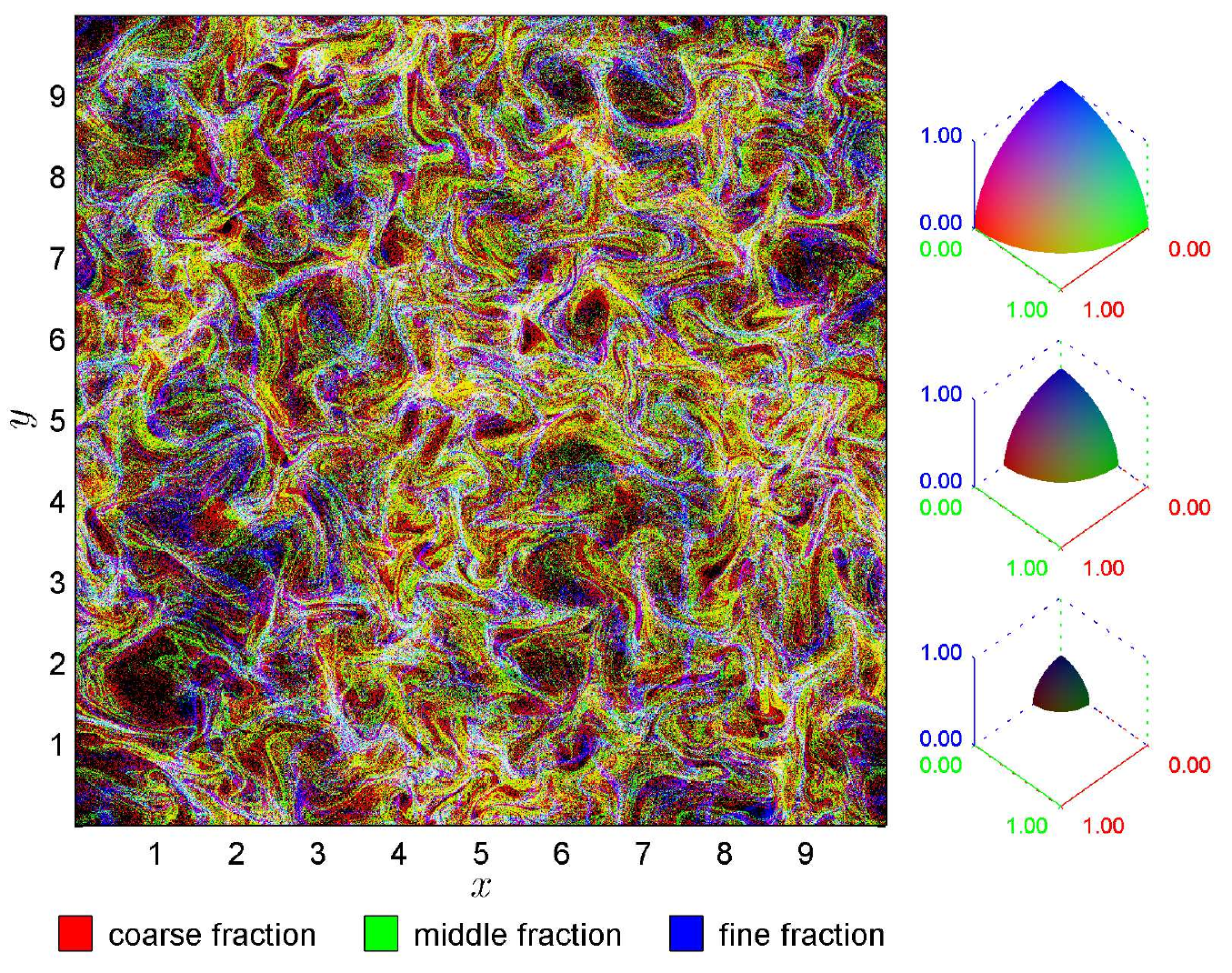}}\\
\end{minipage}
\caption{
Synthetic image of polydisperse dust, consisting of three fractions, for the same time point as in Fig.~\ref {Fig2}. The total number of particles in each fraction is the same. The red color corresponds to the dominance of the concentration of large dust particles, green---medium-sized dust, blue---small dust particles. All other colors correspond to a mixture of fractions in various proportions (for a detailed explanation of the procedure for calculating the color, see section \ref{visualization}).
}
\label{Fig3}
\end{figure}

As a result, the walls of a multiscale cellular structure of the flow with a characteristic minimum size on the order of the pump turbulence scale in the numerical model of 1 pc are clearly distinguishable on the combined image constructed according to the proposed `tricolor 'technique  (Figure~\ref{Fig3}). The vortex cell centers are either almost empty, which corresponds to dark tones, or are filled with fine dust displayed in blue. Larger particles are carried out by vortical motions to the periphery of the vortices.

Considering the fact that in the interstellar medium the dominant synthesis channel for the $H_2$ molecule is the reaction taking place on dust particles \cite{HollenbachMcKee1979}, it can be expected that in places of increased dust particle concentration, molecular hydrogen is formed more intensively.

Fields of concentrations of dust particles of different sizes visualize vortices of different scales.
The contours of the vortices are most clearly drawn by specks with Stokes numbers $Sk \sim 1 \div 10$. This means that small specks of dust trace a fine vortex structure of turbulence, while large dust particles trace extended vortices. Given the fact that the cellular structure in developed turbulence is hierarchical and multiscale, a fraction with particles of the same size will be allocated different colors on different scales. One needs a higher resolution in the calculations to see this effect.

\begin{figure}[h]
\begin{minipage}[ht]{1.0\linewidth}
\center{\includegraphics[width=1.0\linewidth]{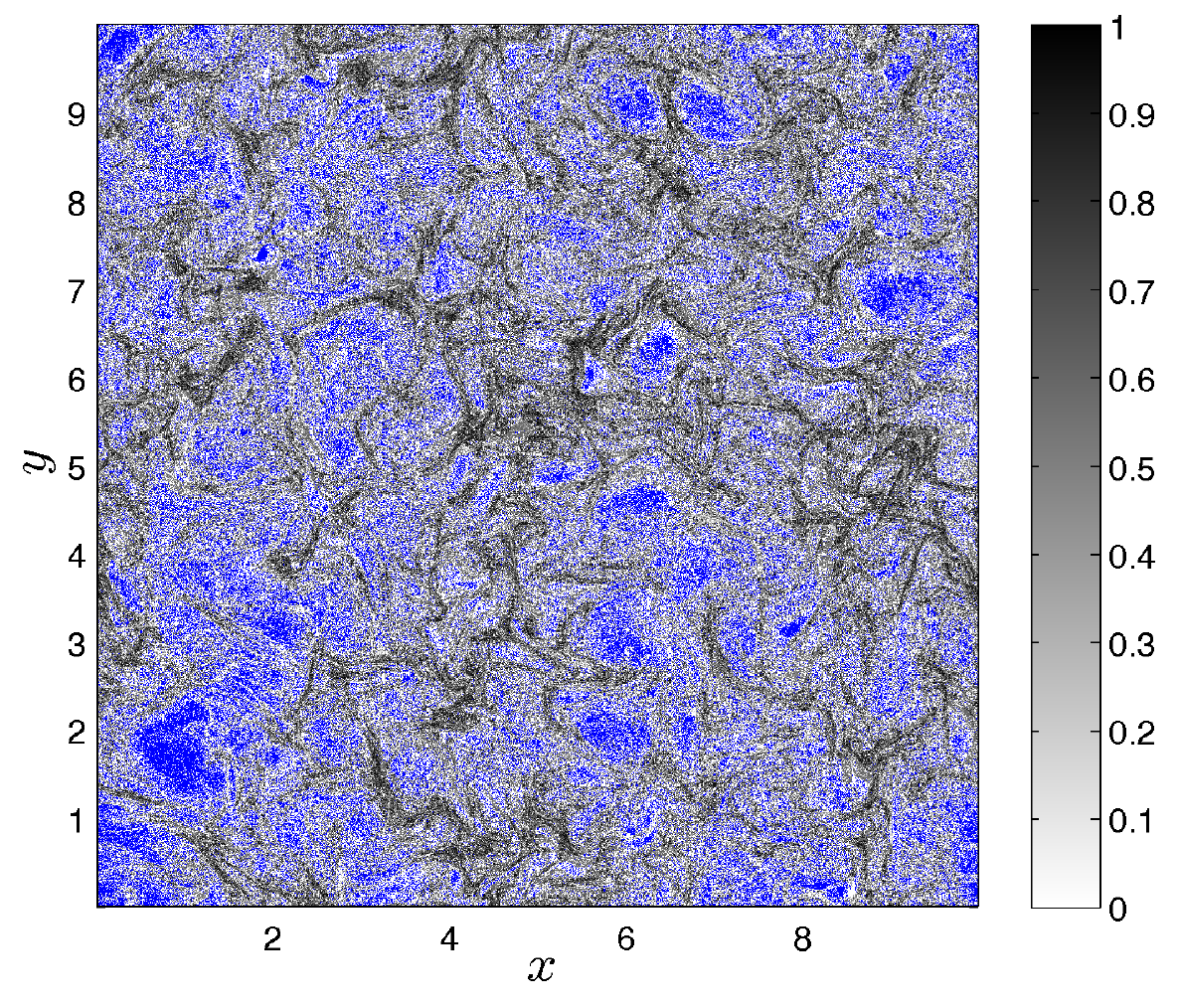}} \\ 
\end{minipage}
\caption{
The entropy distribution of a mixture of three dust particles at the same time as in Fig.~\ref{Fig2} and \ref{Fig3}. The black color corresponds to a homogeneous polydisperse mixture, white is a monodisperse mixture, the areas free of dust particles are highlighted in blue.
}
\label{FigEntropyMix}
\end{figure}

The spatial distribution of the entropy of the dust mixture is illustrated in Figure~\ref{FigEntropyMix}. Light tones in the figure correspond to low values of entropy, close to zero. In these cells the mixture is monodisperse or close to it. The dark cells correspond to values of entropy close to the maximum possible. In these cells, the dust is polydisperse and equidistributed  in fractions. Cells in which there is no dust at all are highlighted in blue. It can be seen that the areas of maximum concentration of particles correspond to the regions of maximum polydispersity, and in the peripheral regions of the bunches bordering the voids, the mixture is monodisperse. In these places, obviously, are concentrated mainly large particles.

\section{Spatial Variations of Dust in a Gas-Dust Flow Through a Spiral Arm of a Disk Galaxy}\label{GUV}

Segregation of dust particles by size is observed in the Galaxy as in the local interstellar medium \cite{Gon_local}, in the galactic disk \cite{Zas,Gon_arm} and in the gas-dust galactic halo \cite{Gon_out}. According to \cite{Zas}, there is a gradient in the distribution of dust particles in size with the distance from the galactic center, with large particles larger than a micron predominating near the nucleus of the Galaxy, when both small particles, smaller than a micron, are at a distance of the order of the radius of the solar orbit. In the paper \cite{Gon_arm} the opposite effect is seen. Large particles begin to dominate the periphery of the Galaxy, i.e., the observed trend of particle size distribution is reversed at distances greater than the radius of the Sun's orbit. The author assumes that the spiral density waves play the role of a large-scale dust segregator in a disk. As a result of segregation, large dust particles concentrate mainly on the outer side of the spiral arms toward the center of the Galaxy, and light dust dominates on the inside.

Numerical modeling makes it possible to demonstrate in detail how and why there is a segregation of dust particles in size in spiral arms,

Let us consider a fragment of gas-dust flow $2 \times 2$ kpc in the equatorial plane of the galaxy near the spiral galactic arm in the region outside the corotation radius.
The coordinates $x$ and $y$ are given in the interval $[- L, L]$, where $L$ is the half-thickness of the gravitational well, which is chosen equal to 1 kpc.

\begin{figure}[ht]
\begin{minipage}[ht]{1.0\linewidth}
\centering{\includegraphics[width=0.4\linewidth]{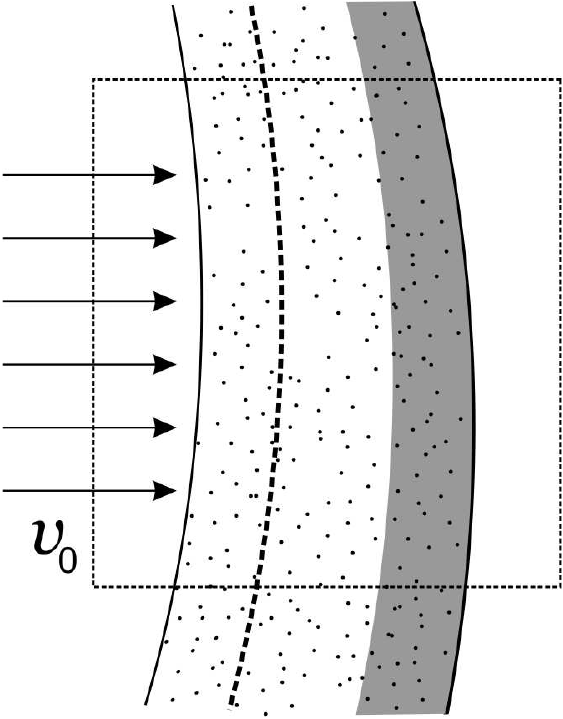}}\\
\end{minipage}
\caption{
Schematic representation of a supersonic flow of gas and dust flowing onto the spiral arm of the galaxy. The fragment of the arm, considered in the calculations, is marked with dashed lines. In the first approximation, the curvature of the arm in the computational domain is neglected. The dashed line indicates the position of the front of the galactic shock wave.
}
\label{Fig0}
\end{figure}

The gas is assumed to flow from the left edge of the computational domain with a supersonic velocity, which causes the formation of a shock front on the front side of the arm with respect to the incoming flow (see Figure~\ref{Fig0}. The Mach number $M_{ini}$ of the input stream is set to 2.

The gas flow in the calculations is turbulized by perturbing the gas parameters at the entrance to the calculated region using a technique similar to that described in the previous section. If the gas were not turbulized, the shock front would be located on the front side of the well at a definite point $x=x_{sh}$, prescribed by the one-dimensional adiabatic GSW theory \cite{KovLev}. The shock front in an unsteady turbulent flow oscillates with respect to its mean position $x_{sh}\approx 0.65 L$ with a characteristic amplitude $0.15 L$.

In the model under consideration, dust particles are created behind a shock front at a distance of $\sim 200 \div 500$ pc from it. The choice of such a distance is due to the characteristic time of formation and evolution of young stars that are born behind the front of the GSW and are carried by the gas flow. In the physics of galaxies, GSW is regarded as a trigger for the process of star formation in interstellar gas, and massive young stars whose rapid evolution is accompanied by the intense stellar wind outflow or explosions like supernovae, the major dust suppliers in the spiral arms of galaxies \cite{Jones06}.

In addition to the frictional force, the force of gravity of the spiral arm and the pressure force of the radiation of young stars act on the dust particles.

The gravitational force can be found from the potential $\Phi$ of the gravitational field of the spiral arm
\begin{equation}\label{force_gravity}
\mathbf{f}_{g} = -\nabla\Phi.
\end{equation}
The cross-sectional dimension of the arm ($\sim 1$ kpc) is much smaller than the radius of curvature of its longitudinal bending ($\sim 10$ kpc), so the spiral arm was approximated in calculations as having an infinite radius of curvature. The gravitational potential well was set homogeneously along the $y$ axis and inhomogeneously along the $x$ axis. In the frame of reference connected with the arm, the potential of its gravitational field was given in the form \cite{Roberts69}
\begin{equation}\label{potential_well}
    \Phi(\mathbf{r})=\Phi(x)=-\frac{\Phi_0}{2}\left(\cos\left(\frac{\pi x}{L}\right)+1\right).
\end{equation}
The depth of the potential well $\Phi_0$ was set equal to $-0.7 c^2_{s, ini}$, where the speed of sound in the gas at the entrance to the computational domain $c_{s, ini}$ was assumed equal to 10 km/s.

The main sources of radiation that accelerate or retard dust particles in a spiral arm are the young massive stars of the $O$ and $B$ classes that are born behind the GSW front due to the star formation process initiated by the GSW. Born stars are carried away by a stream of gas, they line up in stellar rows along the shock front and are displaced relative to the front some distance downstream. For the set of individual point stellar radiation sources, the characteristic distance between which is small compared to the scale of the problem, it is natural to adopt the approximation of a narrow quasi-one-dimensional source layer extended along the $y$ axis.
In this approximation, the light sources are assumed to be distributed along an infinitely thin line in the direction of the axis $y$, located perpendicular to the $x$ axis, intersecting it at the point $x_r$, with a homogeneous luminosity density $\chi$ (luminosity per unit length) .
The value $x_r$ was given in numerical modeling as a time constant and was fixed equal to $-0.3 L$ (Figure~\ref{Fig0}).

On the source line, we select an elementary source of length $dy$, located at the point with coordinates $(x_r, y)$. The source acts by its radiation on a grain of dust at a point with coordinates $(x_d, y_d)$ with a specific force equal in magnitude to
\cite{Jones06, V83}:
\begin{equation}\label{light_pressure_diff}
  df_{r} =\frac{\pi a^2}{m_d} <Q_{pr}> \left( \frac{\chi d y}{4\pi (\Delta x^2+ \Delta y^2) c }\right).
\end{equation}
Here, $\Delta x=x_d-x_r$ and $\Delta y=y_d-y$ are the particle coordinates in the reference frame associated with the source. The total force of the radiation pressure is composed of the forces acting from each of the elementary sources on the source line:
\begin{equation}\label{light_pressure}
  f_{r} =\int_{-\infty}^{\infty}{\frac{df_r}{d y} d y} = \frac{3}{16 a\rho_d} <Q_{pr}> \frac{\chi}{c \Delta x}.
\end{equation}

According to calculations \cite{V83} for spherical graphite particles illuminated by radiation of $B$ stars, the parameter $<Q_{pr}>$ can be approximately specified as a piecewise-continuous dependence
\begin{equation}\label{Q_pr}
\begin{split}
<Q_{pr}> = 0.16          \left(\frac{a}{0.1 \mu m}\right), \quad & 0.01 \le a \le 0.1 \mu m, \\
<Q_{pr}> = 0.16 \phantom{\left(\frac{a}{0.1 \mu m}\right)}, \quad & 0.1 \le a \le 1 \mu m.
\end{split}
\end{equation}

The linear density of luminosity $\chi$ can be estimated as the ratio of the average luminosity $L_*$ of a bright young star to the characteristic distance between bright young stars $l_*$. If the luminosity of the star of the $B$ -class $L_* \sim 10^5 L_{\odot}$ is characteristic, and the characteristic distance $l_* =5.8$ pc, then for the total radiation strength we obtain
\begin{equation}\label{light_pressure_final}
  f_{r} = \left\{
            \begin{array}{ll}
\left(\frac{L_*}{10^5 L_{\odot}}\right) \left(\frac{5.8 pc}{l_*}\right)
\frac{1}{\Delta x}, & \hbox{$0.01 \le a \le 0.1 \mu m$;} \\
\left(\frac{0.1 \mu m}{a}\right)
\left(\frac{L_*}{10^5 L_{\odot}}\right) \left(\frac{5.8 pc}{l_*}\right)
\frac{1}{\Delta x}, & \hbox{$0.1 \le a \le 1 \mu m$.}
            \end{array}
          \right.
\end{equation}

Stellar rows are simultaneously considered as sources of dust production.
Dusts are injected into the calculated region from a randomly chosen point on the source line with some random initial velocities $\mathbf{v}_{d, ini}$.
The distribution density of the birth points of the dust particles is constant and is the same on the entire line $x=x_r$. The direction of the velocity vector is chosen as being uniformly distributed in the interval $[0, 2 \pi]$, and the density distribution of the absolute value of the velocity is given by a non-zero value on the interval $(0, 2\sqrt{2} c_{s, ini}]$, which linearly grows with $v_{d, ini}$. Outside this interval the distribution density is zero.

\begin{figure}[h]
\begin{minipage}[h]{0.485\linewidth}
\center{\includegraphics[width=1\linewidth]{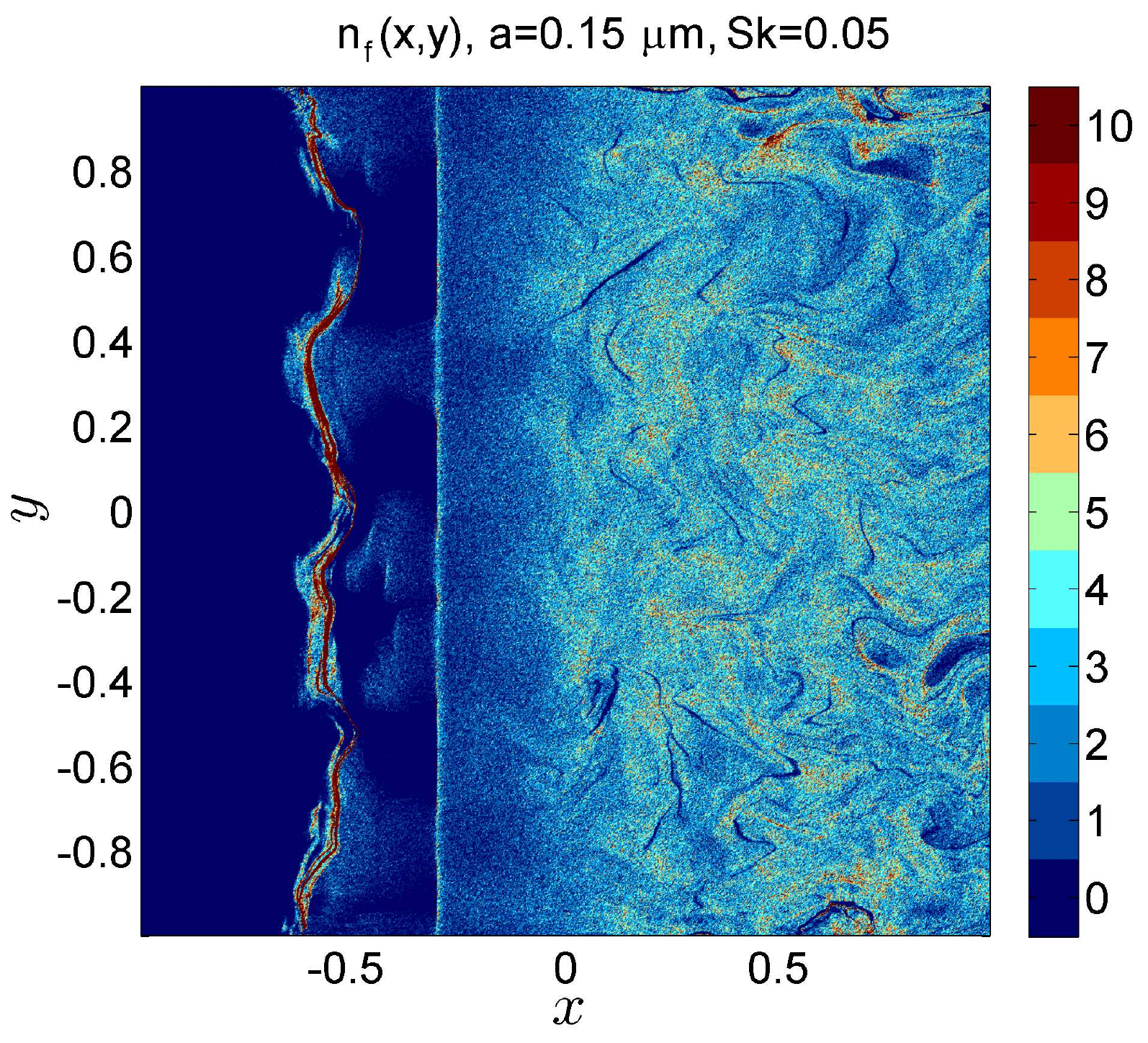}} a) \\
\end{minipage}
\hfill
\begin{minipage}[h]{0.485\linewidth}
\center{\includegraphics[width=1\linewidth]{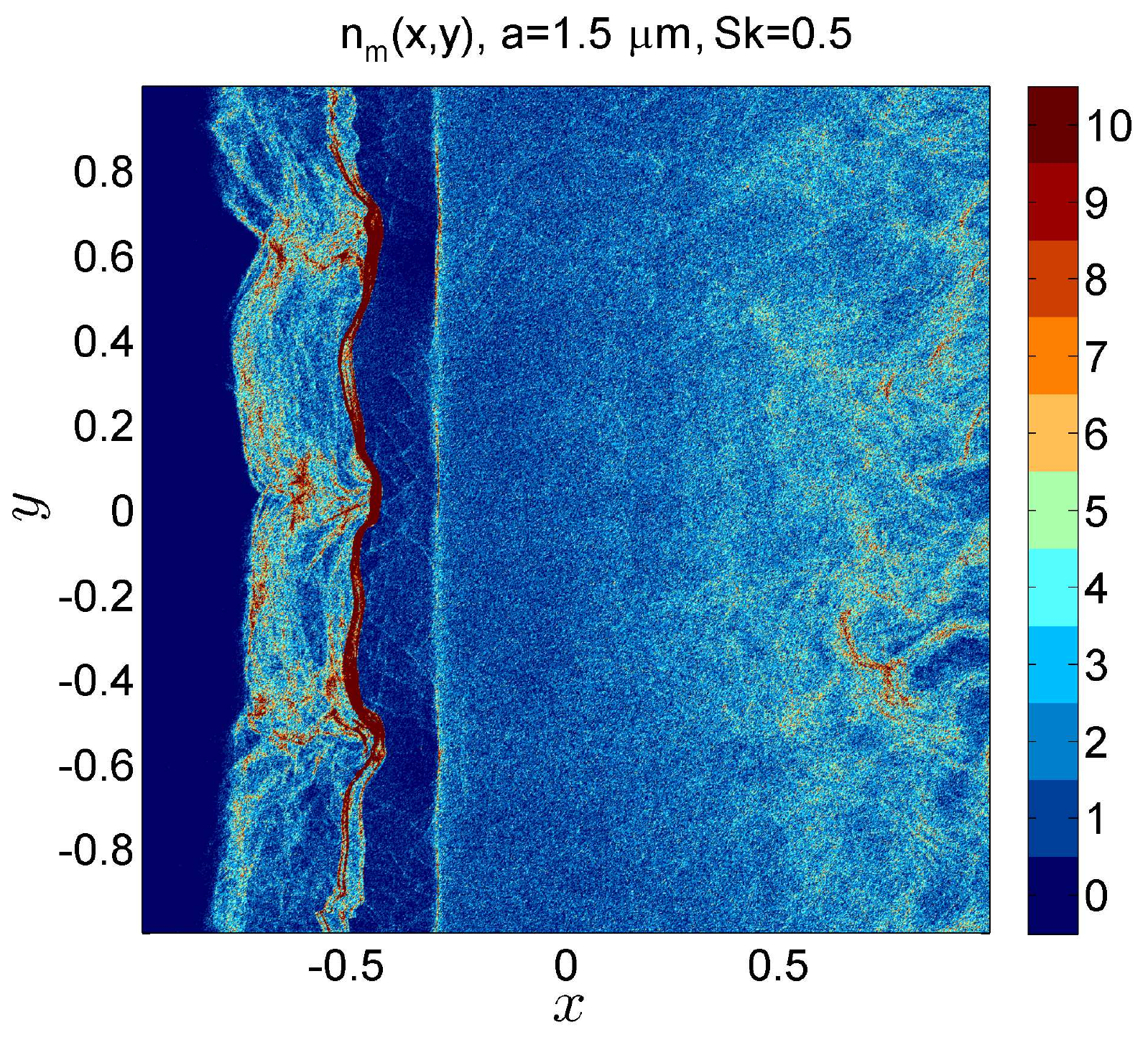}} \\b)
\end{minipage}
\vfill
\begin{minipage}[h]{0.485\linewidth}
\center{\includegraphics[width=1\linewidth]{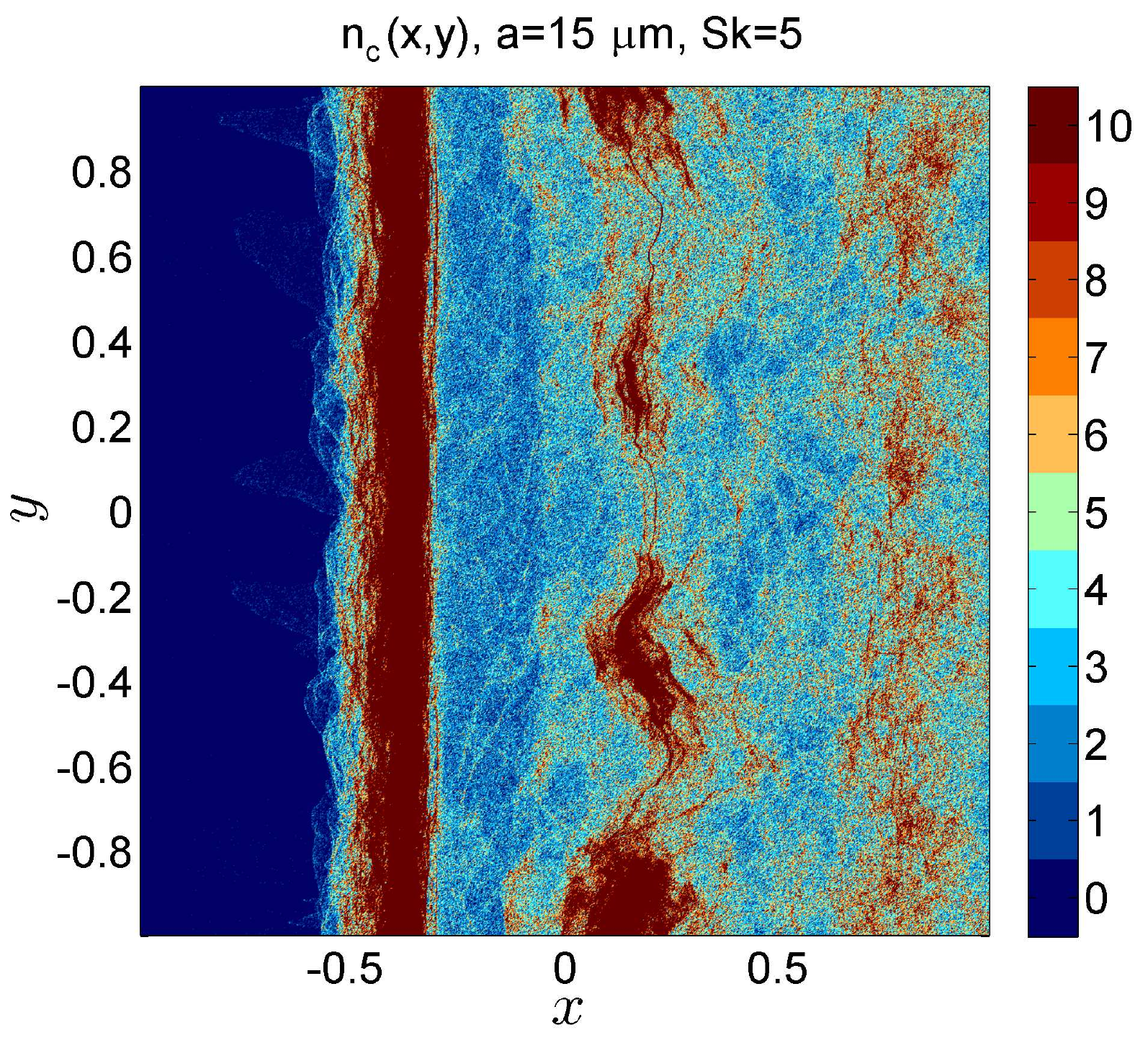}} c) \\
\end{minipage}
\hfill
\begin{minipage}[h]{0.485\linewidth}
\center{\includegraphics[width=1\linewidth]{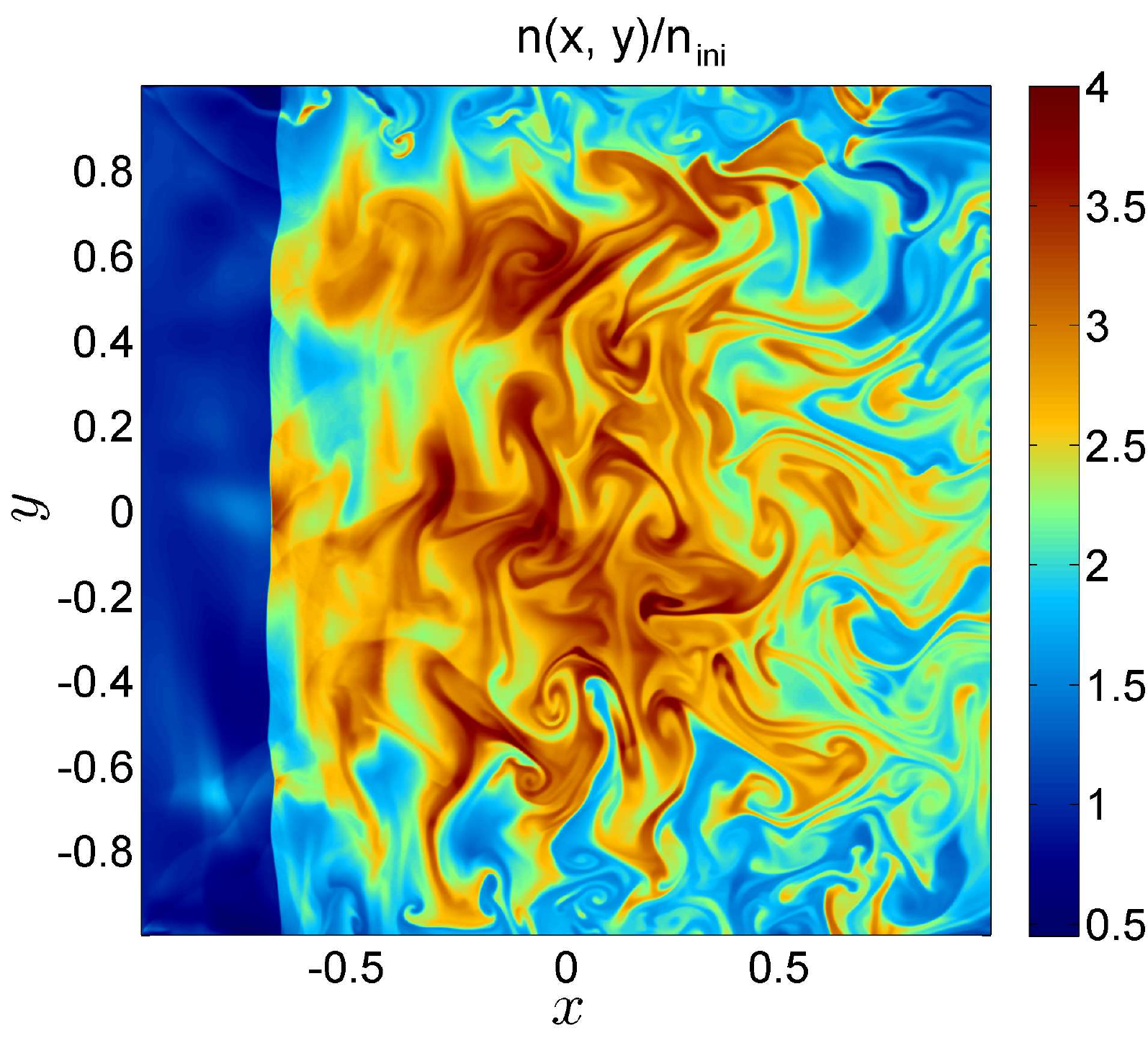}} d) \\
\end{minipage}
\caption{
Distributions of concentrations of dust fractions at the same time after the establishment of dynamic equilibrium in the numerical model of gas-dust flow through the spiral arm of the galaxy. The gravitational well is located in the region $-1 \le x \le 1$, is variable along the axis $x$ and is homogeneous along $y$. The gas-dust mixture moves from left to right. Distributions are given for dust particles of different sizes: (a) fine dust particles with a radius of $a=0.15$ $\mu$m, (b) dust particles of medium size with a radius of $a=1.5$ $\mu$m, (c) large specks, $a=15$ $\mu$m.
The corresponding values of the Stokes numbers $Sk$ are indicated in the headings to the figures. Figure (d) shows the distribution of the gas density at the same time.
The density is normalized to the density value at the entrance to the computational area.
}
\label{Fig4}
\end{figure}

The dust emitted by the sources against the gas flow is held by the radiation pressure, partially or completely, depending on the size of the dust particles, and forms on the front side of the spiral arm the dust accumulations along the arm (see Figure~\ref{Fig4}). This exactly reproduces the observed phenomena in practically all spiral galaxies dust lanes along the edges of the spiral arms. In those areas where lanes are concentrated, there is a force balance. For lanes on the leading edge of the arm, the forces of gravitation and friction that push the dust particles along the flow are compensated for by the counteraction of the radiation pressure force.

The turbulence of the flow destroys the regular flow structure along the streamline. It causes fluctuations in the position of the shock front and forms corrugation of lanes and their subdivision into separate clusters. The wavering shock front carries away the dust particles. Small particles with Stokes numbers $Sk=0.05$ quickly relax to dynamic equilibrium. The lanes made up of small dust particles take the outline of the shock front (Figure~\ref{Fig4} a). In the same figure, a vertical whitish line is clearly visible, which determines the position of the line of light and dust sources.

Larger particles are weakly accelerated by the shock front and react more slowly to frequent front fluctuations. As a result, dust lanes containing dust particles of medium size are more extendeded along the flow (Figure~\ref{Fig4} b). In their structure, a broad relatively sparse band is traced, which is formed due to the fact that the dust particles, once attracted by the vibrating shock front over long distances upstream, do not have time to quickly return to the point where the front is at the moment. A dense thin strip on the inner edge of the lane, close to the source, is formed from dust particles concentrated in the forefront region and succeeded in balancing to an equilibrium position.  A shock front can no longer carry away these particles and carry them out against the stream.

The largest dust particles react poorly to the impact of the shock front (Figure~\ref{Fig4} c). Their main mass is concentrated in the interval $x \approx -0.5 \div -0.3 $ in the region between the source line and the first turning point of their trajectory when they are ejected by the source in the direction opposite the flow.

On the outer side of the gravitational well with respect to the flowing stream, the conditions for the implementation of the balance of forces can also be realized. Here the force of gravity acts as a restoring force, and the action of force on radiation pressure is negligible. For small and medium sized dust particles, friction is large, so they are carried away by the flow and carried out of the arm, while large dust particles, for which the friction force is weaker than the gravitational force, settle on the back side of the well near the $x \approx 0.2$ line, performing damped aperiodic oscillations in the vicinity of this equilibrium position. On (Figure~\ref{Fig4} c), dense lanes near the equilibrium position are clearly visible as well as in the vicinity of the first right turning point for $x \approx 0.7$.

The above dynamic model of the gas-dust flow of the interstellar medium in the spiral arm of a flat galaxy can explain the formation of the observed powerful dust lanes on the front edge of the spiral arm and weak dust lanes on the rear edge of the arm.

\begin{figure}[h]
\begin{minipage}[ht]{1.0\linewidth}
\center{\includegraphics[width=1.0\linewidth]{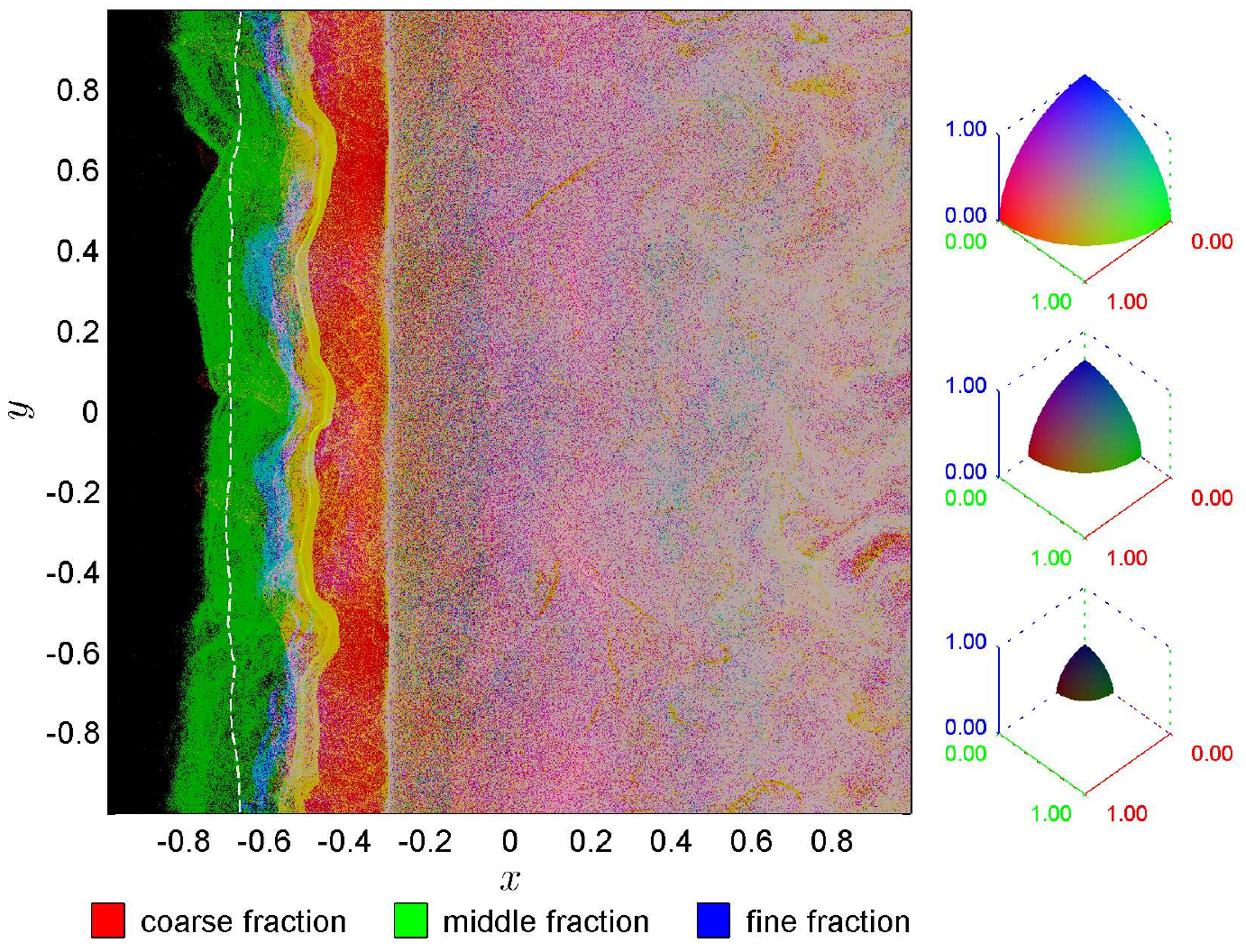}}
\caption{
Synthetic image of a two-dimensional pattern of the flow of polydisperse dust in a turbulent gas-dust flow through a gravitational potential well of a spiral sleeve for the same time point as in Fig.~\ref {Fig4}. The dust component contains three fractions, large dust particles (red color), dust particles of medium size (green color) and small particles (blue). To the right of the figure are the sections of the cubic palette of colors by spheres of different radii, see the caption to Fig.~\ref {Fig3}. A detailed explanation of the construction of a color scheme for a mixture with arbitrary proportions between the concentrations is given in the section \ref{visualization}. The white dashed line indicates the position of the front of the galactic shock wave for a given time. A thin whitish line at $x \approx 0.3$ corresponds to the position of a combined light and dust source.
}
\label{Fig5}
\end{minipage}
\end{figure}

The combined image (Figure~\ref{Fig5}), constructed according to the technique `tricolor ', allows directly observe the result of segregation of dust particles in size.

The black area to the left of the well corresponds to the almost complete absence of dust particles in the inflowing stream, with the exception of single red dots marking large dust particles. Then follows a layered region of about 0.5 kpc in width, in which the dust particles are clearly separated in size, and the color of each layer allows us to draw conclusions about the type of population. Medium-sized dusts form a wide zone of green along the shock front with a width of 0.1-0.2 kpc, in which individual yellow filaments are visible, formed by a small amount of large dust particles.
The transition to the blue hue marks a narrow band at $x \sim 0.6$ kpc, in which light dust particles are collected. They move to the left under the influence of radiation pressure and decelerate in the gas flow. The next yellow band is formed in the region $x \sim 0.5$ kpc when a thick lane of medium-sized dust particles and the left boundary of a band of large dust particles occurs. The latter one in turn creates a red layer with filaments of yellow and purple hues marking the chains of dusts of medium and small dimensions carried along the stream.

With the help of the applied `tricolor' visualization algorithm, it is possible to clearly visualize the segregation of particles by the dimensions inside the spiral arm. On the front edge of the arm between the shock front and the source lines, bands of different colors are distinctly visible, the positions of which reflect the predominance of one or another fraction in the given region. Fine dust is visible in the form of a thin bluish strip. Dusts of medium size, distributed evenly in these area, form a broad green band. The yellow band corresponds to the position of the mixture of particles, where the medium and large dust particles are contained in equal parts. Near the source line and on the back edge of the arm behind the source line, shades of red predominate, which corresponds to the predominance of large particles.

It can be clearly seen that large grains of dust tend to accumulate on the back side of the spiral arm, creating a region that is predominantly red, which corresponds to a thick dust band at the rear edge of the arm. Fine dust concentrates on the front, this area is visible as a thin bluish strip. Dust of medium size is distributed more or less evenly. In general, this picture is in good agreement with the segregation of dust particles observed in spiral arms of various sizes.

\section{Conclusions}

In this paper, we present a numerical algorithm for visualizing the spatial distribution of polydisperse dust in an inhomogeneous flow of a gas-dust mixture. The technology of decomposition of an arbitrary color in three basic colors is used, which allows us to instantly reproduce in one figure the polydisperse dust content at each point of space in the computational area.
With the help of the proposed technique, one can visualize the degree of polydispersity of the mixture by graphical representation of the spatial distribution of the entropy of a mixture of dust particles, which allows us to observe the effects of spatial separation of dust fractions (segregation of dust particles in size). For the flows in which dust clusters are formed, the proposed technique makes it possible to reveal the ability of dust particles from different fractions to clustering. The examples of numerical modeling of the flow of a mixture of gas and polydisperse dust in a turbulent cloud and in a shock wave demonstrate the usefulness of the visualization procedure, since it helps to identify those areas of flow where dust particles of various sizes are mainly concentrated.
Spatial segregation of dust particles by size can serve as a marker for explaining the features of complex hydrodynamic flows of a multiphase medium.
Thus, the proposed technique can potentially serve as a means of diagnosing the turbulent velocity and density field in a gas-dust medium.

\section{Patents}
The tricolor technique for visualization of the hydrodynamic jumps by using the rate-of-strain tensor is protected by the Russian Federation Certificate [Korolev V.V. The program for the hydrodynamic flow  structure analysis by using the rate-of-strain tensor. –- RF Certificate of the Computer Program State Registration N 2012615602, 20.06.2012].
\vspace{6pt}

{\bf Acknowledgments:} This study was supported by the grant 16-31-00466-mol\_a from the Russian Foundation for Basic Research.



\end{document}